\documentclass[amstex,epsf,amssymb,usenatbib]{mn2e}
\usepackage{epsfig}
\usepackage{amsmath,amssymb}
\newcommand{\Msolar}{\mbox{\,$\rm M_{\odot}$}}        

\title[G-mode pulsations in sdB stars: a test of stellar opacity]{Gravity-mode pulsations in subdwarf B stars: a critical test of
stellar opacity}

\author[C.S. Jeffery \& H. Saio]
       {C.S. Jeffery\thanks{E-mail: csj@arm.ac.uk}
\& H. Saio\thanks{E-mail: saio@astr.tohoku.ac.jp} \\
Armagh Observatory, College Hill, Armagh BT61 9DG, Northern Ireland\\
Astronomical Institute, School of Science, Tohoku University, Sendai 980-8578, Japan\\
}

\date{Accepted .....
      Received ..... ;
      in original form .....}

\pagerange{\pageref{firstpage}--\pageref{lastpage}}
\pubyear{2006}

\begin{document}

\maketitle

\label{firstpage}

\begin{abstract}
\noindent The identification of non-radial g-mode
 oscillations as the cause of variability in cool subdwarf B stars 
 (PG1716 variables) has been frustrated by a 5\,000\,K discrepancy
 between the observed and theoretical blue edge of the instability
 domain \citep{Fon03}. A major component in the solution to this
 problem has been identified by (a) using
 updated OP 
 instead of OPAL opacities and (b) considering an enhancement of
 nickel, in addition to that of iron, in the driving zone. The reason
 for this success is that, in OP, the ``Fe-bump'' contributions 
 from iron and nickel occur at higher temperatures than in OPAL. As well
 as  pointing to a solution of an important problem 
 in stellar pulsation theory, 
 this result provides a critical test for stellar opacities
 and the atomic physics used to compute them. 
\end{abstract}

\begin{keywords}
 atomic processes, radiative transfer, stars: interiors, 
 stars: oscillations, stars: horizontal branch, stars: early-type
\end{keywords}

\section{Introduction}              
\label{intro}

\citet{Jef06} recently reviewed the
excitation of pulsations in certain low-mass stars by so-called ``Fe-bump
instability''. This instability is caused by  a significant
contribution to stellar opacity from M-shell electrons in iron-group elements 
at temperatures around 200\,000\,K \citep{OPAL92, OP95}.
It provides the driving mechanism for radial pulsations in certain extreme
helium stars \citep{Sai93} and p-mode oscillations in hot subdwarf B stars
(EC14026 variables: after prototype EC\,14026--2647 $\equiv$ V361\,Hya) 
\citep{Cha96,Cha97,Kil97}. 

The Fe-bump opacity mechanism is effective in these stars because of
an increase in the contrast between opacity due to iron-group elements 
and opacity from other sources. In the case
of extreme helium stars, the background hydrogen opacity is
suppressed. In the case of EC14026 stars, \citet{Jef06} demonstrated that,
for radial and non-radial p-mode oscillations, the observed boundaries
of the instability strip can be explained only by increasing the
iron abundance ($X_{\rm Fe}$) without increasing the heavy element 
abundance ($Z$) as a whole. It was already well known that 
radiative forces act differentially on the ions 
in the stellar envelope such that substantial chemical gradients 
are established over a diffusion time scale $\sim 10^5$y
\citep{Mic89}. 
The consequent levitation and accumulation of iron in layers at 
around  200\,000\,K \citep{Cha95} enhances the Fe-bump opacity
sufficiently to excite pulsations in about 10\% of sdB stars within
the EC14026 instability zone \citep{Cha01}. 
 
The  discovery of oscillations with periods
of a few hours in many cool sdB stars (PG1716 variables: after
PG\,1716+426) has presented a challenge
to stellar pulsation theory \citep{Gre03}. 
While the radiatively-driven diffusion of iron can still 
operate in these stars, p-modes were reported to be 
stable in the chemically stratified models 
\citep{Cha01}. On the other hand, non-radial g-modes of high radial order 
($k\geq10$) and high spherical degree ($l\geq3$) were found to be
unstable, but only in models of stars cooler 
than those in which variability has been detected \citep{Fon03}. 

\citet{Jef06} found that, for homogeneous models of blue horizontal
branch stars, non-radial g-modes of high radial order ($k\geq10$) 
and low spherical degree ($l\ge2$) could be excited even without 
any enhancement of the iron abundance, but only in a narrow range 
of effective temperature around 18\,000\,K. 
The width of the g-mode instability strip and the number of excited
modes increases substantially with
increasing iron abundance, but \citet{Jef06} were unable to obtain
unstable g-modes for $T_{\rm eff}>24\,000$\,K. Even these modes were
only obtained with iron enhanced by a
factor of twenty over a base metallicity $Z_0=0.02$ and for
$l=3$. This theoretical g-mode blue edge is some 5\,000 K cooler than
the hottest known PG1716 variables, Balloon 090100001 and HS\,0702+6043, 
which have $T_{\rm eff}\sim29\,500$K \citep{Ore05,Sch06}. Efforts 
to reduce this difference by varying the helium and 
iron abundances, and the overall metallicity, were unsuccessful. 

Subsequently the authors have considered what other
mechanisms might contribute to the instability of g-modes in PG1716 stars.  
This paper describes the surprising results that the explanation lies
in the atomic physics used to compute the stellar opacities (section
2) and that the theoretical blue-edge can be reconciled with observation
by considering the individual r\^oles of other iron-group elements
including nickel (section 3). 
The consequences for theoretical models of stellar 
interiors and the atomic processes to be found therein are briefly
considered (section 4), before conclusions are drawn (section 5).

\begin{figure*}
\begin{center}
\epsfig{file=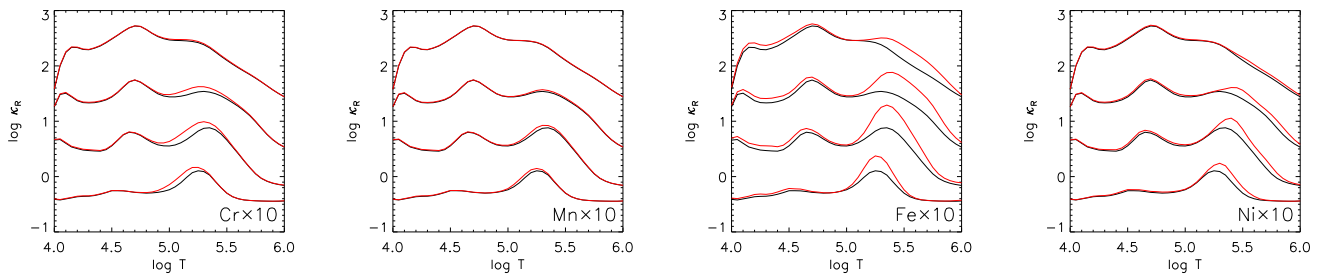,width=18cm,angle=0}
\caption[OP elemental opacities]{
Comparison of $\log(\kappa_{\rm R})$ from OP for $X=0.70, Z_0=0.02$ (black)
with the same mixture with Cr, Mn, Fe, and Ni (left to right) 
enhanced selectively by a factor 10 (red). 
Curves are given for $\log(R) = -2, -3, -4$ and $-6$ (top to bottom), 
where $R=\rho/T_6^3$, $\rho$ is mass desnity in g\,cm$^{-3}$ and $T_6$ is
$10^6 \times T$ with $T$ in K. 
}
\label{op_elements}
\end{center}
\end{figure*}

\begin{figure*}
\begin{center}
\epsfig{file=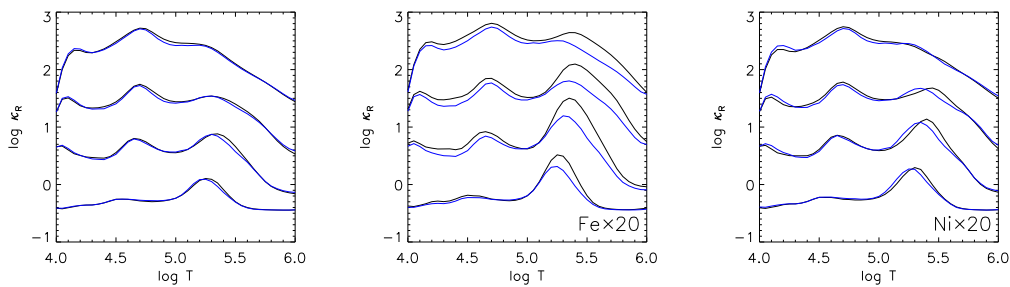,width=14cm,angle=0}
\caption[OP vs OPAL opacities]{
Comparison of $\log(\kappa_{\rm R})$ from OP (black) with that from OPAL 
(blue) for $X=0.70, Z_0=0.02$ (left) and with Fe (center) and Ni (right) 
enhanced by a factor 20. Otherwise as Fig.~\ref{op_elements}
}
\label{op_opal}
\end{center}
\end{figure*}

\begin{table}
\caption{Assumed fraction of iron-group elements within the 
heavy element fraction $Z_0$: OPAL = \citet{Gre93}, OP = S92 of \citet{Sea94}.  }
\label{t_mix}
\begin{center}
\begin{tabular}{llll}
El & $Z$ & \multicolumn{2}{c}{$Z_{\rm el}/Z_0$} \\
   &     &  OPAL  & OP  \\[1mm]
Cr & 24 & 0.000329 & 0.000324\\
Mn & 25 & 0.000170 & 0.00017 \\
Fe & 26 & 0.021877 & 0.02244 \\
Ni & 28 & 0.001293 & 0.00123 \\
\end{tabular}
\end{center}
\end{table}

\section{Stellar opacity}

The motivation for this work is the general result that the position of an
instability strip, {\it i.e.} the $T_{\rm eff}$ range within which a star
can be expected to pulsate, is dictated by the temperature of the
opacity peak which provides the driving. In their original work, 
\citet{Jef06} considered variations in only helium ($Y$), all metals 
together ($Z_0$), and
iron ($Z_{\rm Fe}$) 
-- the latter being the dominant element within the 
iron group by a large factor. 
However, if radiative acceleration acts on iron,
it will also operate on other elements, particularly iron-group
elements with electronic structures similar to that of iron. 

Our first question was, therefore, whether an increased contribution
of elements {\it other} than iron might alter the position ({\it i.e.}
temperature) of the ``Fe-bump''. This was addressed 
by obtaining tables of updated Rosseland mean opacities \citep{Bad05}
from the Opacity Project webserver for mixtures with $X=0.7$,
$Z_0=0.02$, 
but with elements chromium, manganese, iron and nickel individually
enhanced by a factor ($f$) of ten. Comparing these in the 
vicinity of the  ``Fe-bump'', it was evident that there
are significant differences in the bump temperature for each
element (Fig.~\ref{op_elements}). 
Considering the relative abundances of these elements
(Table~\ref{t_mix}) and that we had simply multiplied by ten 
the default fraction
of each element  (as prescribed by the respective webservers), 
it was also apparent that nickel 
and chromium are highly effective absorbers. Significantly,
the bump temperature of nickel is markedly higher than that of
iron, whilst those of manganese and chromium are similar or lower. 
Therefore, we set out to investigate whether a nickel overabundance 
might drive g-mode pulsations at a higher $T_{\rm eff}$ than iron.

Since our stellar structure code is designed to work
with OPAL opacity tables, a new set of tables was obtained from the 
OPAL webserver with nickel enhanced by a factor $f_{\rm Ni}=20$, 
and other elements normal  ($X=0.70, Z_0=0.02$). A pulsation stability 
analysis was carried out for a series of zero-age horizontal-branch 
models, as described by \citet{Jef06}. These
new opacities had a negligible effect on the stability results, being
essentially the same as for the mixture with $f_{\rm Ni}=1$ and
$f_{\rm Fe}=20$. 

The reason became apparent when the OPAL and OP opacities were
compared. The peak temperatures of the bumps are significantly
different in the two sets \citep[cf. Figs. 1 and 10,][]{Sea04}.
The Fe-bump occurs at a
higher temperature in OP than in OPAL. The OPAL Ni-bump is at
practically the same temperature as the OPAL Fe-bump 
(Fig.~\ref{op_opal}). It would be necessary to make the
pulsation calculations using OP data. 

This was achieved by obtaining single tables of Rosseland mean opacity 
from the OP webserver for mixtures
$X=0.70,0.35,0.10,0.03,0.0, Z_0=0.02$, and reformatting these to match 
the OPAL tables using the OP utility  {\tt opfit} \citep{Sea93}.
Type 2 OPAL tables were obtained from the OPAL webserver for 
the same standard mixtures, but including the 16 tables with 
enhanced C and O used to compute evolved stars as required by our codes.
Both sets of tables were computed  
with elemental enhancements $f_{\rm Fe,Ni}=(1,1)$; $(20,1)$;
$(1,20)$; and $(10,10)$. 

To use the OP tables within our codes, the OPAL
tables were read in first. The OP tables were then read in to
replace the corresponding tables with no enhanced C and O. 
This works because the interpolations required for 
modelling the H-rich layers of ZAHB stars make no use of the 
tables with enhanced C and O. 

\begin{table}
\caption{Blue-edge ($\log T_{\rm eff}/$K) of 
  instability strip for g-modes in blue horizontal branch stars}
\label{t_blueedge}
\begin{center}
\begin{tabular}{lc lll lll}
$f_{\rm Fe,Ni}$ 
        &     & OP   &      &      & OPAL &      &      \\ 
        & $l$ & 1    & 2    & 3    & 1    & 2    & 3    \\[2mm]
1,1     &     & 4.26 & 4.31 & 4.34 & --   & 4.26 & 4.29 \\
20,1    &     & 4.36 & 4.41 & 4.43 & 4.32 & 4.36 & 4.39 \\
1,20    &     & 4.39 & 4.43 & 4.45 & 4.31 & 4.35 & 4.38 \\
10,10   &     & 4.38 & 4.43 & 4.45 & 4.33 & 4.37 & 4.40 \\
\end{tabular}
\end{center}
\end{table}

\begin{figure*}
\begin{center}
\epsfig{file=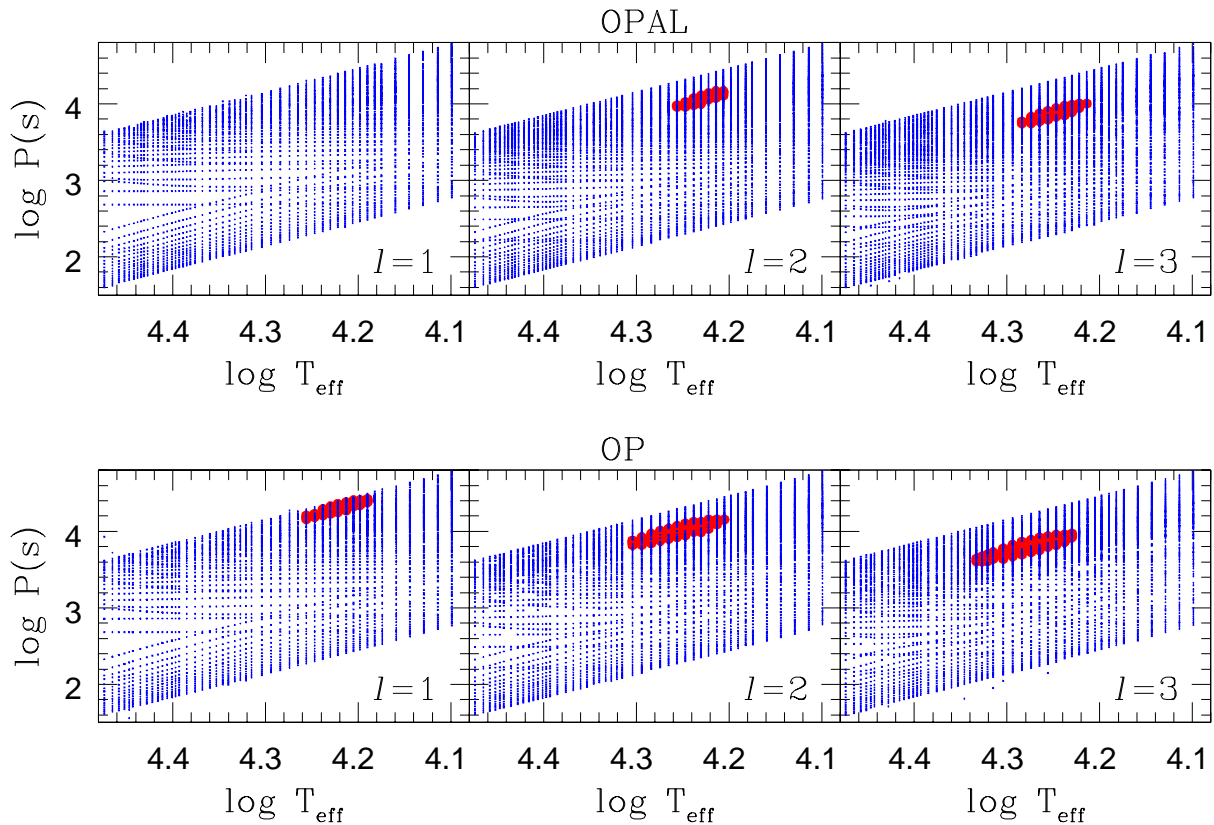,width=16cm,angle=0}
\caption[Non-radial stability analysis: $f_{\rm Fe,Ni}=1,1$]
{Periods of modes due to Fe-bump instability for
  ZAHB stars with $M_{\rm c}=0.486\Msolar$, $X=0.75$, $Z_0=0.02$, and
  $M_{\rm e}=$ increasing from left to right. The top row
  shows models computed with OPAL opacities, the bottom row
  with OP opacities. Stable modes are marked as
  (blue) dots, unstable modes are marked as filled (red) circles.
}
\label{g_normal}
\end{center}
\end{figure*}

\begin{figure*}
\begin{center}
\epsfig{file=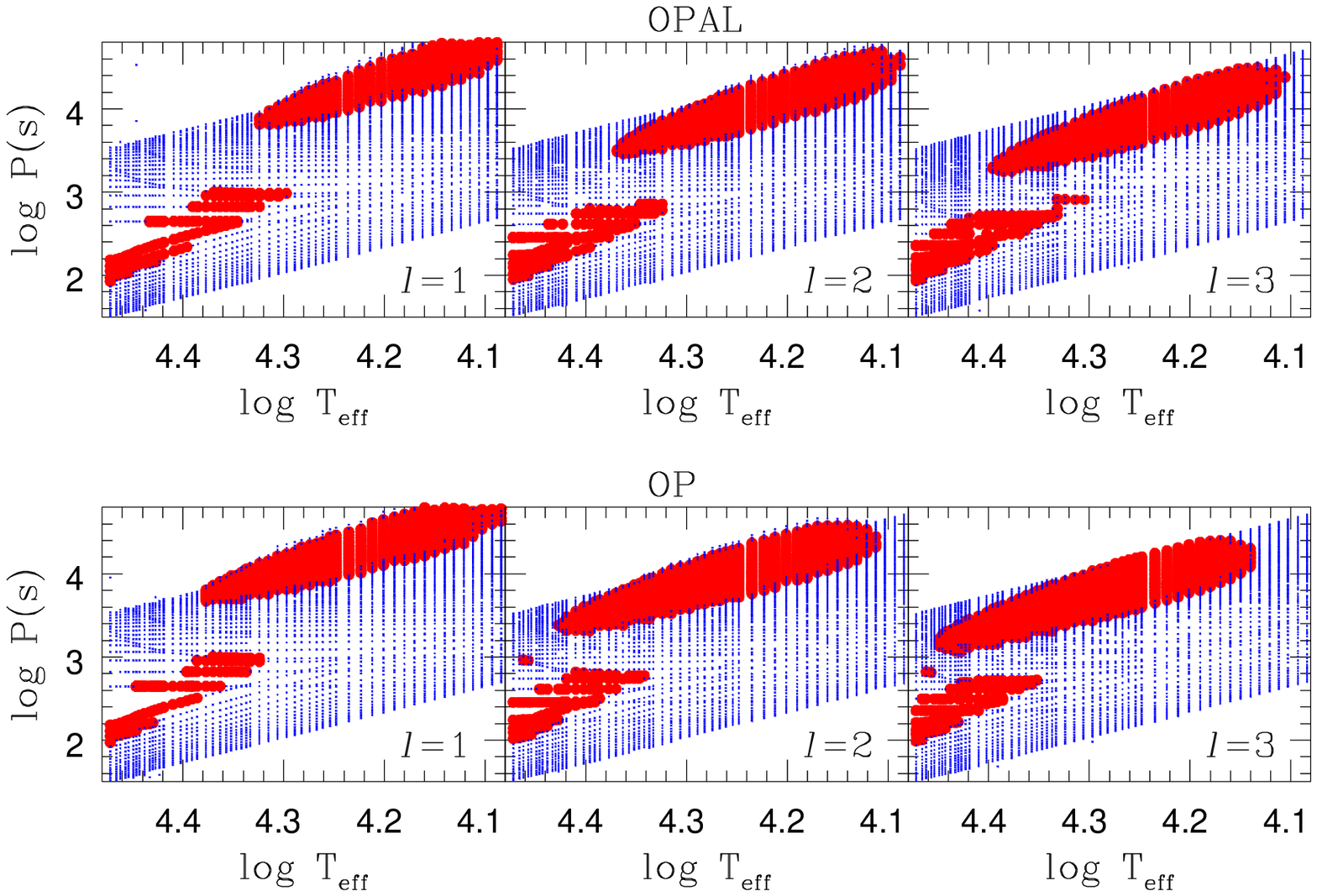,width=16cm,angle=0}
\caption[Non-radial stability analysis: $f_{\rm Fe,Ni}=10,10$]
{As Fig.~\ref{g_normal}, for models with 
  iron and nickel both enhanced by a factor $10$. PG1716-type 
  variability has been observed for stars in the range 
  $4.32 \lesssim \log T_{\rm eff}/{\rm K} \lesssim 4.47$ 
  with periods $3.41 \lesssim \log P/{\rm s} \lesssim 3.95$.}

\label{g_fe10ni10}
\end{center}
\end{figure*}

\section{g-modes in blue horizontal-branch stars}

As previously \citep{Fon03,Jef06}, we have assumed that the PG1716
variables are extreme horizontal branch stars and have 
constructed series of ``zero-age'' horizontal-branch (ZAHB) models
having a helium-burning core with a mass of $M_{\rm c}$
and with a hydrogen-rich envelope with mass $M_{\rm e}$ ranging from
 $3\times10^{-6}$ to 0.034 \Msolar. 
Here we have considered only the case of $M_{\rm c} = 0.486\Msolar$.
The surface layers are characterized by mass
fractions of hydrogen $X=0.75$ and base metallicity $Z_0=0.02$. 
with iron and nickel enhanced by the pairs of  
$f_{\rm Fe},f_{\rm Ni}$ given above. 
\citet{Jef06} discuss variations with $X$ and $Z_0$, and $M_{\rm c}$. 
 In particular, varying $M_{\rm c}$ has no effect on the stability of
 g-modes. They also found  the only significant effect of
  evolution is that g-modes cease to be excited when the core 
  becomes radiative at the  end of core He-burning. Of course 
  pulsation periods $P$ increase with radius $R$. 
  Since $P \propto R^{3/2}$ 
  or surface gravity $g^{-3/4}$, normal horizontal-branch
  evolution produces a period increase of $\sim 0.2$ dex in all
  modes.

The high-temperature end of this ZAHB sequence corresponds with the locus of
the short-period EC14026 variables. PG1716 variables are to be found
at lower temperatures, 
with $21\,000 \lesssim T_{\rm eff}/{\rm K} \lesssim 29\,000$ 
\citep{Fon06,Ran06}. The
lower limit probably does not represent a formal red edge; 
horizontal branch stars at these temperatures are very scarce.

We have tested the stability of each of our
horizontal-branch models for {\it non-radial} p- and 
g-modes with spherical degree $l=1,\ldots,4$. 
The frequency range considered is 
$0.2 \le \omega \le 20$, where $\omega$ is the angular frequency of pulsation
normalized by $\sqrt{GM/R^3}$ with $G$ being the gravitational constant. 

Fig.~\ref{g_normal} compares the results for OP and OPAL with normal
composition. With OP, unstable modes are obtained even for $l=1$. In
general, the number and the temperature range of unstable modes is
increased with OP opacities. 
Fig.~\ref{g_fe10ni10} compares the results for OP and OPAL with
$f_{\rm Fe,Ni}=(10,10)$. For modes with $l=3$, g-modes are
excited up to $T_{\rm eff}/{\rm K}\lesssim 28\,000$ in OP models, but
only to  $T_{\rm eff}/{\rm K}\lesssim 25\,000$ in OPAL models.
Although less likely to be observable, they are excited up to  
$T_{\rm eff}/{\rm K}\lesssim 29\,500$ for OP models with $l=4$. In
contrast, the position of the models on the HR diagram hardly changes
with the choice of opacity table. 

Table~\ref{t_blueedge} compares the theoretical blue-edge of the 
g-mode instability strip for different enhancements of iron and
nickel, for OP and OPAL models and for the three most observable modes,
$l=1,2$ and 3. 

It is clear that the use of OP opacities combined with the
inclusion of excess nickel can shift the theoretical blue-edge of 
the g-mode instability strip significantly closer to the observed blue
edge. This prompts us to believe that \citet{Fon03} correctly
identified the oscillations in PG1716 variables as opacity-driven
g-modes, and that the discrepancy in the predicted and observed blue 
edges can be solved by invoking more accurate atomic physics in the 
calculation of stellar opacity, as well as considering atomic species
other than iron. 

The basic elements of this solution are that (i) the opacity peaks due to
iron and nickel occur at a higher temperature in OP than in OPAL, 
(ii) the OP nickel peak occurs at a higher temperature than 
the OP iron peak, (iii) at these temperatures nickel is a  
more efficient absorber than iron (per ion), and (iv) 
radiative acceleration will force it to accumulate in the Ni-bump 
region in exactly the same way as iron. 

\section{Atomic physics}

It should, perhaps, have been no surprise that the introduction of
OP opacities would shift the instability blue edge. A more modest
shift in the blue edge for high-order g-modes in slowly pulsating B
stars was obtained by \citet{Pam99}. 
Since then, extensive work on 
the contribution of inner-shell electrons to the OP
calculations, has brought the OP opacities into ``much closer
agreement with those from OPAL'' \citep{Bad04,Bad05}. 

Directed at obtaining good atomic data for very large numbers of
atomic transitions, OPAL uses single-configuration wavefunctions with
one-electron orbitals \citep{OPAL92}. 
OP originally obtained most of its atomic data
using the R-matrix method to evaluate multi-electron wavefunctions, 
resulting in more accurate atomic data, but for fewer and
simpler transitions \citep{OP95,OP97}. Prompted by substantial differences
between the first OP results and OPAL, primarily at high 
temperature and densities, the introduction of
configuration-interaction codes enabled OP to include many more
bound states and radiative transition probabilities, as
well as auto-ionization probabilities and
photo-ionization cross-sections \citep{Sea05}.  

For many astrophysical situations, OP and OPAL opacities
appear to be in closer agreement with one another ($<5\%$) \citep{Sea05} 
than with observation. For example, helioseismology results currently
require an increase of $\sim20\%$ in opacity below the base of the
solar convective zone \citep{Bah05}.  While the results
presented here do not directly suggest a solution to the solar opacity
problem, they {\it do} indicate that improvements in
the treatment of atomic processes used in the calculation of stellar
opacities can still have substantial consequences for calculations of
pulsation stability. These are especially important when amplified by 
radiatively-driven diffusion of elements into layers 
of high specific opacity, as in the outer layers of many
early-type stars. In turn, these will impact on elemental
stratification, on stellar radii, on boundaries of pulsational
stability, and on pulsation periods derived for use in
asteroseismology.

\section{Conclusion}                
\label{conclusion}

We have analyzed the stability of models of hot zero-age 
horizontal-branch stars to nonradial g-mode oscillations. We 
have tested and confirmed the hypothesis that elements other 
than iron can raise the temperature of the so-called ``Fe-bump'' in 
opacity at around 200\,000\,K. In doing so we have discovered that,
within this ``Fe-bump'', recently updated OP opacity tables provide 
more opacity at significantly higher temperatures than the OPAL
tables. The higher temperature Fe- and Ni-bumps
in the OP tables have been combined with a tenfold enhancement of both 
nickel and iron abundances in the driving zone. This provides sufficient 
opacity at the right temperatures to shift the theoretical g-mode blue-edge 
to within 1000\,K of the observed blue-edge for PG1716 stars. 
Given
the approximate nature of our own calculations and the likely remaining
uncertainties in the atomic physics, we conclude that, by 
modifying the driving mechanism 
identified by \citet{Fon03}, the discrepany in the predicted and 
observed blue edges  for PG1716 variables can be solved.

In reaching this conclusion, we have
demonstrated that g-mode pulsations in PG1716 stars provide a
critical test of stellar opacity, substantially preferring the updated
OP calculations to OPAL. Future work should include an investigation 
of the use of OP opacities in models for EC14026 variables and 
OP radiative accelerations in self-consistent models of chemical 
diffusion in extreme horizontal-branch stars,  as well as the influence
of important opacity providers other than iron and nickel
(e.g. chromium, manganese, vanadium). Detailed work
for asteroseismological analyses will need to consider a wide range of
additional physics, including the effects of evolution and 
time-dependent diffusion theory and core mass on predicted periods.

\section*{Acknowledgment}

Travel support for this collaborative project was provided through 
PPARC grant PPA/G/S/2002/00546. This work has made extensive use of
of the OP and OPAL webservers.

\end{document}